# Title: Single shot ultrafast all optical magnetization switching of ferromagnetic Co/Pt multilayers

**Authors:** Jon Gorchon[1,2]†, Charles-Henri Lambert[2,]†, Yang Yang[3]†, Akshay Pattabi[2], Richard B. Wilson[4], Sayeef Salahuddin[1,2], Jeffrey Bokor[1,2].

**Affiliations:**

[1] Lawrence Berkeley National Laboratory, 1 Cyclotron Road, Berkeley, CA 94720, USA.
[2] Department of Electrical Engineering and Computer Sciences, University of California, Berkeley, CA 94720, USA.
[3] Department of Materials Science and Engineering, University of California, Berkeley, CA 94720, USA.
[4] Department of Mechanical Engineering and Materials Science and Engineering Program, University of California, Riverside, CA 92521, USA.

*Correspondence to: jgorchon@lbl.gov, jbokor@berkeley.edu

†Denotes equal contribution.

**Abstract**: A single femto-second optical pulse can fully reverse the magnetization of a film within picoseconds. Such fast operation hugely increases the range of application of magnetic devices. However, so far, this type of ultrafast switching has been restricted to ferri-magnetic GdFeCo films. In contrast, all optical switching of ferro-magnetic films require multiple pulses, thereby being slower and less energy efficient. Here, we demonstrate magnetization switching induced by a single laser pulse in various ferromagnetic Co/Pt multilayers grown on GdFeCo, by exploiting the exchange coupling between the two magnetic films. Table-top depth-sensitive time-resolved magneto-optical experiments show that the Co/Pt magnetization switches within 7 picoseconds. This coupling approach will allow ultrafast control of a variety of magnetic films, critical for applications.

**Main Text:**

In a number of recent experiments[1–8], it has been shown that femtosecond laser pulses can control magnetization on picosecond timescales, which is at least an order of magnitude faster compared to conventional magnetization dynamics. Among these demonstrations, the material system, GdFeCo ferrimagnetic films, is particularly interesting because deterministic toggle-switching of the magnetic order is possible without any external magnetic field. This phenomenon is often referred to as all optical switching (AOS). To date, GdFeCo is the only material system where such deterministic switching is observed. When extended to ferromagnetic systems, which are of greater interest in many technological applications, only a partial effect can be achieved, which in turn requires repeated laser pulses for full switching[9–11]. Such repeated pulsing is not only energy hungry, it also negates the speed of AOS. To address this problem, we have developed a method that allows for full reversal of the magnetization of ferromagnetic materials on a picosecond time-scale from a single laser pulse. We demonstrate that in exchange-coupled layers of Co/Pt and GdFeCo, single shot, switching of the ferromagnetic Co/Pt layer is achieved within 7 picoseconds after irradiation by a femtosecond laser pulse. This approach will greatly expand the range of materials and applications for ultrafast magnetic switching.

Our study focuses on a set of samples whose structure consists of a perpendicularly magnetized ferromagnetic Co/Pt multilayer grown (see the supplementary materials) on a perpendicular GdFeCo layer (see Fig. 1a). The aim of these structures is to exploit the coupling between the two layers in order to extend the single-shot AOS capabilities of GdFeCo to the ferromagnet. We chose Co/Pt as the ferromagnet for its strong perpendicular magnetic anisotropy even when grown on top of a non-textured film such as GdFeCo and for the possibility of increasing its thickness (number of repeats) all while keeping the perpendicular anisotropy. The stacks were characterized by performing hysteresis loops with a magneto optical Kerr effect (MOKE) microscope (see the supplementary materials) and an out-of-plane magnetic field $H_\perp$. Hysteresis loops are shown in Fig.1b. Four remnant states are present in samples d=4 and 5 nm, and only two in the strongly coupled d=1.5-3 nm. We add a quarter wave plate in the optic path to enable depth-sensitive MOKE [12], which allows us to obtain layer sensitivity, as demonstrated in the loops of Fig.1c. The polarity of the loops allows us to infer the directions of the magnetic moments in the stack (see the supplementary materials). We indicate the direction of the Gd, FeCo and Co/Pt sublattice magnetizations by orange, green and blue arrows respectively. We note that the net moment of the GdFeCo/Co structures is dominated by the FeCo lattice (see the supplementary materials).

To characterize the type of the interlayer coupling of the stacks, we measure magnetic hysteresis loops of the magnetic stacks. We obtained the coupling for samples d=4 and 5 nm by performing minor hysteresis loops, where only the Co/Pt magnetization is switched (blue circles in Fig.1b). The interlayer bias field $H_b$ corresponds to the shift of the minor loop. The GdFeCo and Co/Pt net moments present an AFM coupling for thick spacers, as was found in Refs.[13–15]. An RKKY-type of exchange [13–16] or a dipolar orange peel coupling [13,17] could explain such AFM coupling. However, in Gd dominated samples, $H_b$ has an opposite sign (example shown in the supplementary materials Fig.S3), demonstrating that the coupling does not follow the direction of the net moment, but rather the orientation of the sublattices. In addition, as shown by the magnetic moments depicted in Figs.1b and c, the coupling changes sign and becomes FM as the spacer is made thinner. We thus attribute the net coupling to an RKKY-type of exchange as reported in Refs.[13–16].

We checked the all-optical switching capabilities of all the previously characterized samples by irradiating the Co/Pt side with 70fs (FWHM) linearly polarized laser pulses. Digitally re-colored MOKE micrographs of the AOS are shown in Fig.2a. Refer to Fig.2b for a schematic of the color scheme used for the various magnetic states found in the first row of images in Fig.2a. The samples are initialized via a positive external magnetic field into a state where the GdFeCo and Co/Pt magnetizations are parallel. The magnetic field is then turned off. After the first single laser shot on sample d=5nm we observe two new regions of different contrast, a centered white circle and the surrounding purple ring. Both new regions present opposite anti-parallel alignments of GdFeCo and Co/Pt. In the center area, the laser intensity is above the critical fluence $F_C$ for AOS (blue line in Fig.2.b), which results in the reversal of the GdFeCo magnetization. The Co/Pt magnetization remains in its initial orientation in order to relax the structure into a more stable AFM state. In the surrounding ring, the fluence is below threshold for AOS and GdFeCo does not switch. However, the hot Co/Pt layer does switch in order to relax into the AFM state. There is thus a second threshold $F_C$ for the relaxation of the parallel state to antiparallel (red line in Fig.2.b). For even lower fluences, the heating is insufficient to generate any observable changes. When the sample is shot with a second laser pulse, the magnetization in the center region switches fully, reversing both GdFeCo and Co/Pt moments. The ring region remains stable after the first shot, as it is already in a stable AFM state. We then repeat the experiment for different thicknesses d of the Pt spacer. Sample d=4nm presents the same type of switching as sample d=5nm. For samples d=1.5, 2 and 3 nm, where the coupling is FM, we observe only the switching of the center area, as expected, as no more remnant states are available.

We illuminate the samples repeatedly and obtain reliable toggle switching of both ferromagnetic and ferrimagnetic layers for up to more than 100 pulses. The switching is reproducible in all FM and AFM coupled samples. Therefore, we have consistently demonstrated single-shot AOS in various ferromagnetic films. We emphasize that the pulses used in all experiments are linearly polarized, thereby excluding any type of helicity dependent magneto-optical mechanisms[9,18,19].

The absorbed critical fluences $F_c$ for AOS, for relaxation as well as the estimated equilibrium lattice temperature rises $\Delta T$ (see the supplementary materials) are reported in Fig.2c by black, red and blue points, respectively. We obtain a per pulse temperature rise of $150 \pm 30$ K for all films, irrespective of the type of coupling. The Co/Pt it thus heated to $450 \pm 30$ K, very close to its expected Curie temperature of ~470 K [20]. This indicates that for AOS to be possible the Co/Pt needs to be nearly fully demagnetized. Another possible mechanism to consider for switching are hot electron spin currents between the layers during demagnetization [4]. If present, spin currents should have a different sign for parallel and antiparallel alignments, resulting in different critical fluences. Because $F_c$ for AOS is equal for AFM and FM coupled films, we believe that spin currents are probably not important to the switching mechanism.

The MOKE micrographs in Fig.2.a provide clear evidence of the AOS of Co/Pt, but cannot provide information on whether we are achieving ultrafast control over the exchange interaction, which is the ultimate goal of our study. In order to access the fast magnetization dynamics, we perform depth-sensitive time-resolved MOKE measurements [21] with no external magnetic field. Details of the technique are given in the supplementary materials. For this purpose, we grow a second series of two perpendicularly magnetized samples with an extra Co/Pt repeat (Fig.3a), in order to increase the sensitivity of our MOKE measurements to the Co/Pt layer. A patterned Au coil grown on top of the magnet delivers short intense magnetic field pulses at the laser repetition rate of 54 kHz to repetitively reset the magnetization between laser pulses during pump-probe experiments (see the supplementary materials). The Pt spacer thickness d of the samples was chosen equal to 1.5 and 4 nm, to obtain a strong FM and a weaker AFM coupling, respectively. We confirmed the AOS of the full stack in both samples by MOKE imaging.

We then perform low fluence demagnetization experiments in order to check the differences in demagnetization for a sample in an antiparallel (AP) vs a parallel (P) state. For this purpose, we work on sample d=4 nm which has 4 remnant states. We first select the quarterwave plate angles in order to maximize layer sensitivity to the Co/Pt and GdFeCo moments, as shown in Fig.3b. By working at low fluence we avoid the full reversal of the Co/Pt layer, which has too high of a coercivity (~500 Oe) to be reset by the field delivered by the patterned coil. The results of the demagnetization dynamics are shown in Fig.3c. The demagnetization of both GdFeCo and the Co/Pt peaks at around ~1-2ps of delay time, in agreement with recent reports where 60-70fs laser pumps at high intensities were used[7,22]. The Co/Pt and GdFeCo peak demagnetizations are similar in amplitude for both parallel (P) and antiparallel (AP) cases, but the long timescale (~2-20ps) dynamics of the Co/Pt magnetization are very different. This result is rather surprising, since the same energy is deposited in the Co/Pt film in both P and AP cases, which should result in a similar spin temperature and magnetization at long time delays[4]. We attribute the slower recovery of the magnetization in the parallel case to the intrinsic AFM coupling field of the stack that pulls the Co/Pt against the anisotropy field. If spin currents were relevant during demagnetization, they should be maximized during the fast demagnetization and have opposite signs for P and AP cases. The similar peak-demagnetization amplitudes for both P and AP cases indicate that spin currents do not play a major role.

Finally, we perform time-resolved switching experiments. Unfortunately, these experiments could not be carried on sample d=4 nm because the coercivity of the Co/Pt layer is larger than the amplitude of the reset magnetic field we could generate with the patterned coils. We thus perform the AOS experiments on a thinner d=1.5 nm sample, where both layers are strongly FM coupled and present a single and smaller coercivity (H~150 Oe). We vary the incident average power from 40 to 59 mW.

As shown in Fig.3d, the dynamics of the GdFeCo and Co/Pt layers are quite different. At the lowest power, 40 mW, we only obtain a demagnetization of both GdFeCo and Co/Pt magnetizations. Increasing the power to 53 mW results in the switching of GdFeCo within ~3 ps, during which the Co/Pt demagnetizes nearly completely. The magnetization of Co/Pt recovers during the next 4 ps as the system cools down. Eventually, due to the exchange field induced by the GdFeCo sublattice, the Co/Pt magnetization switches after ~30 ps. At even high power, 59 mW, the Co/Pt demagnetizes first and then grows in the opposite direction, switching in only ~7 ps.

The two-steps of the 7 ps switching event of Co/Pt shown in Fig.3d, - an initial full demagnetization and a subsequent switching - strongly supports the idea that the exchange interaction is responsible for the reversal of the softened (hot) Co/Pt magnetization. The curves at 53 and 59 mW indicate that the fastest switching occurs when the fluence is such that the temperature of the Co/Pt reaches exactly (or very close to) $T_C$, i.e a full demagnetization.

We have shown that, despite using thick (up to 5 nm) metallic Pt spacers to separate a Co/Pt multilayer from a GdFeCo layer, we can still achieve single-shot AOS of the ferromagnetic layer by exploiting the exchange interaction. Moreover, we demonstrated a 7 ps switching time on a sample with a 1.5 nm Pt spacer. This rather general method can be extended to other ferromagnets, ferrimagnets, or antiferromagnets. The utilization of high spin polarization materials coupled to GdFeCo should allow for ultrafast control of magnetic devices with high tunnel magneto resistance ratios[23,24]. In addition, such a method for ultrafast laser writing of magnetic recording media without magnetic fields might be an appealing alternative for heat-assisted magnetic recording technology [25]. Finally, the combination of our approach with recent developments in ultrafast all-electronic switching of magnetism[26] enables additional exciting possibilities for picosecond spintronics.

**Supplementary materials:**

See supplementary materials for extra details on the sample properties, the magneto-optical techniques, as well as optical absorption and temperature rise calculations.

**Acknowledgments:**
This work was primarily supported by the Director, Office of Science, Office of Basic Energy Sciences, Materials Sciences and Engineering Division, of the U.S. Department of Energy under Contract No. DE-AC02-05-CH11231 within the Nonequilibrium Magnetic Materials Program (MSMAG). We also acknowledge support by the National Science Foundation Center for Energy Efficient Electronics Science (materials synthesis, sample fabrication, and experimental equipment).

**Author contributions:**

J.G., C.H.L, Y.Y and J.B devised the experiments. C.H.L grew the samples. Y.Y. did the microfabrication. J.G., C.H.L, Y.Y and A.P performed the experiments. J.G. wrote the manuscript with input from all the authors.

**The authors declare no competing financial interests.**

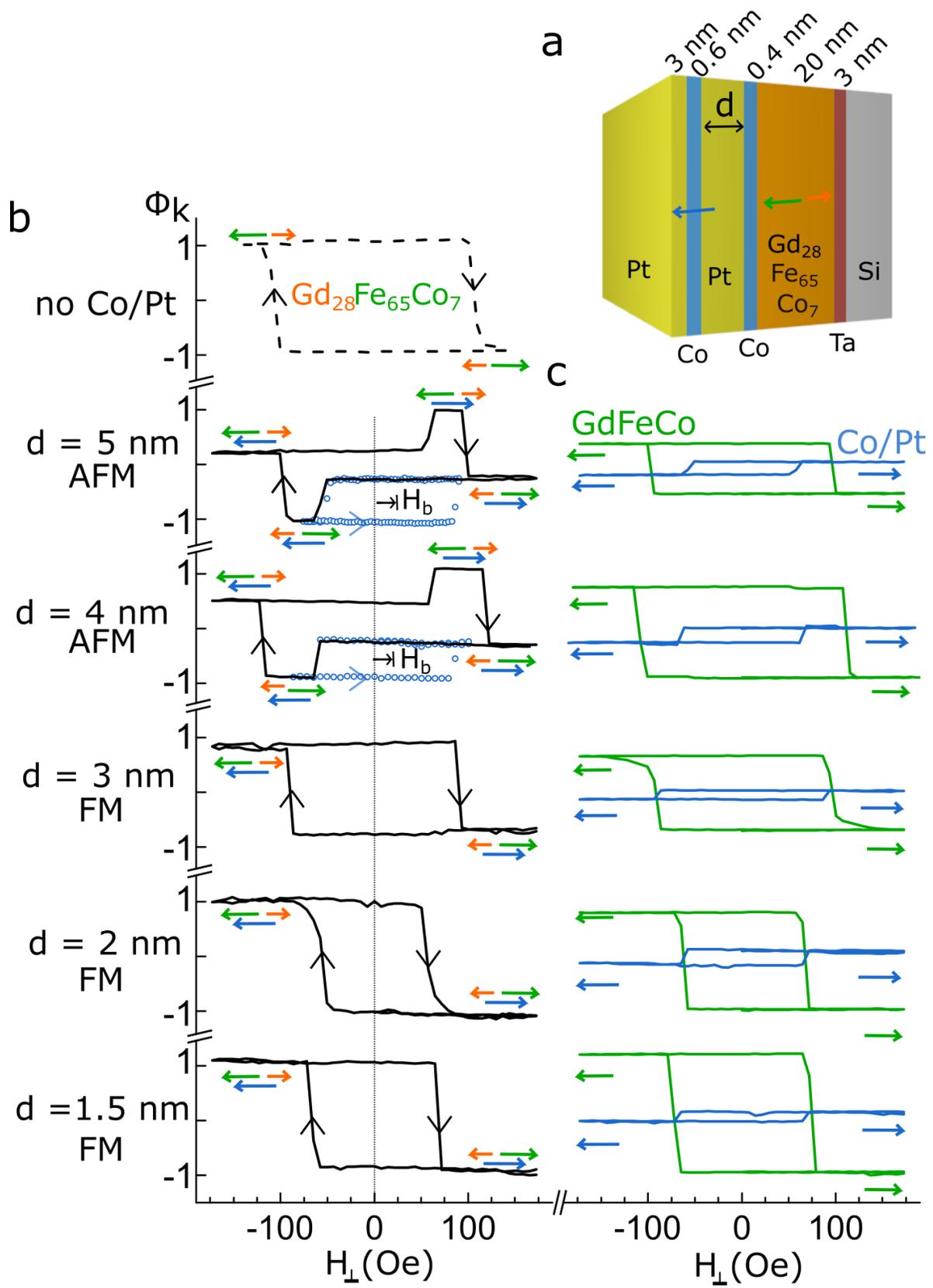

**Fig. 1**. **Sample description and magnetic characterization**. **a,** GdFeCo/Co/Pt(d nm)/Co stack series with Pt spacer of thickness d. **b,** Magnetic hysteresis loops for (top) bare GdFeCo(20 nm)/Ta(3 nm) and the GdFeCo/Co/Pt/Co stacks with different Pt spacer. Orange, green and blue arrows represent Gd, FeCo and Co/Pt magnetizations, respectively. Minor loops switching only the Co/Pt magnetization, presenting a positive exchange bias $H_b$, on samples d=4 and 5 nm. **c,** Depth sensitive MOKE hysteresis loops of the stacks, with maximum sensitivity to GdFeCo (green) and Co/Pt (blue) magnetic lattices.

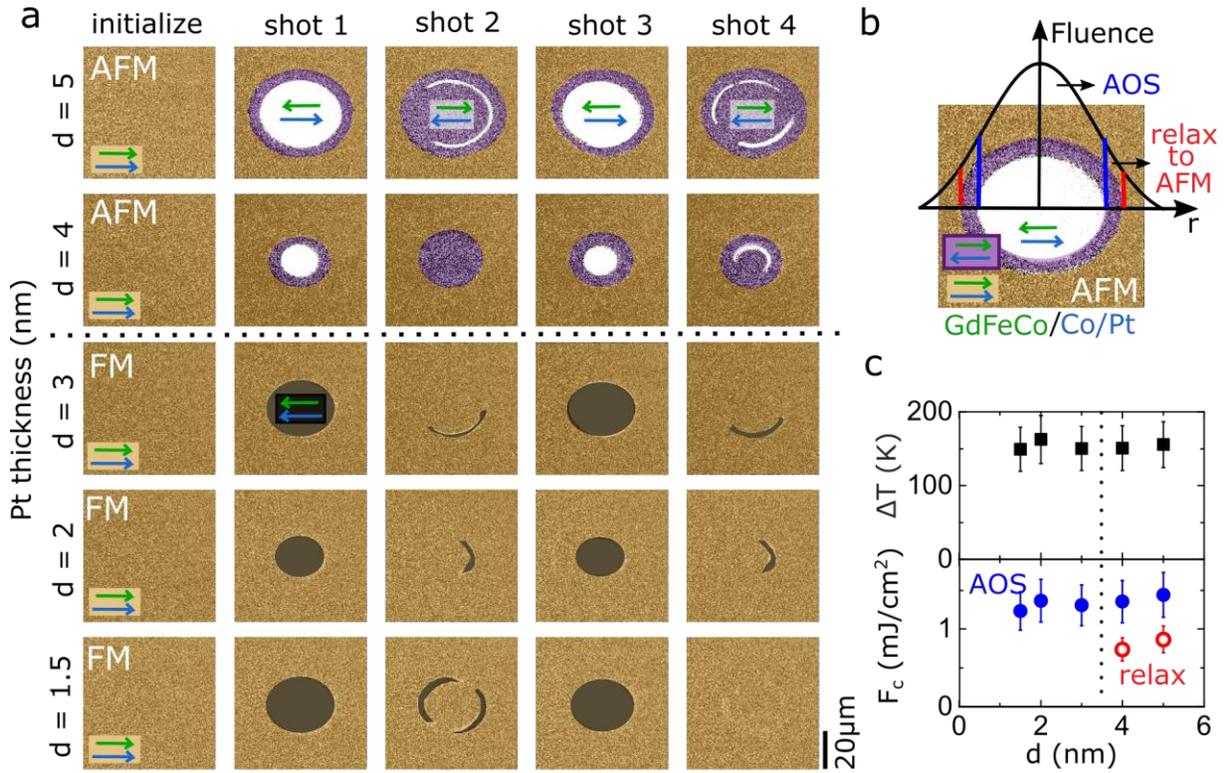

**Fig. 2**. **Single shot all optical switching of both Co/Pt and GdFeCo layers. a,** Digitally re-colored MOKE images of a sequence of AOS events on the GdFeCo/Co/Pt(d nm)/Co stack series presented in Fig.1. Green and blue arrows represent GdFeCo and Co/Pt magnetizations, respectively. **b,** Laser intensity profile and resulting domain configuration on film d=5 nm after

the first laser shot. **c,** Absorbed critical fluence $F_c$ for AOS (blue filled circles), $F_c$ for relaxation of parallel states following the AFM coupling (red empty circles), and estimated peak lattice temperature rises $\Delta T$ (black squares) as a function of the Pt spacer thickness d.

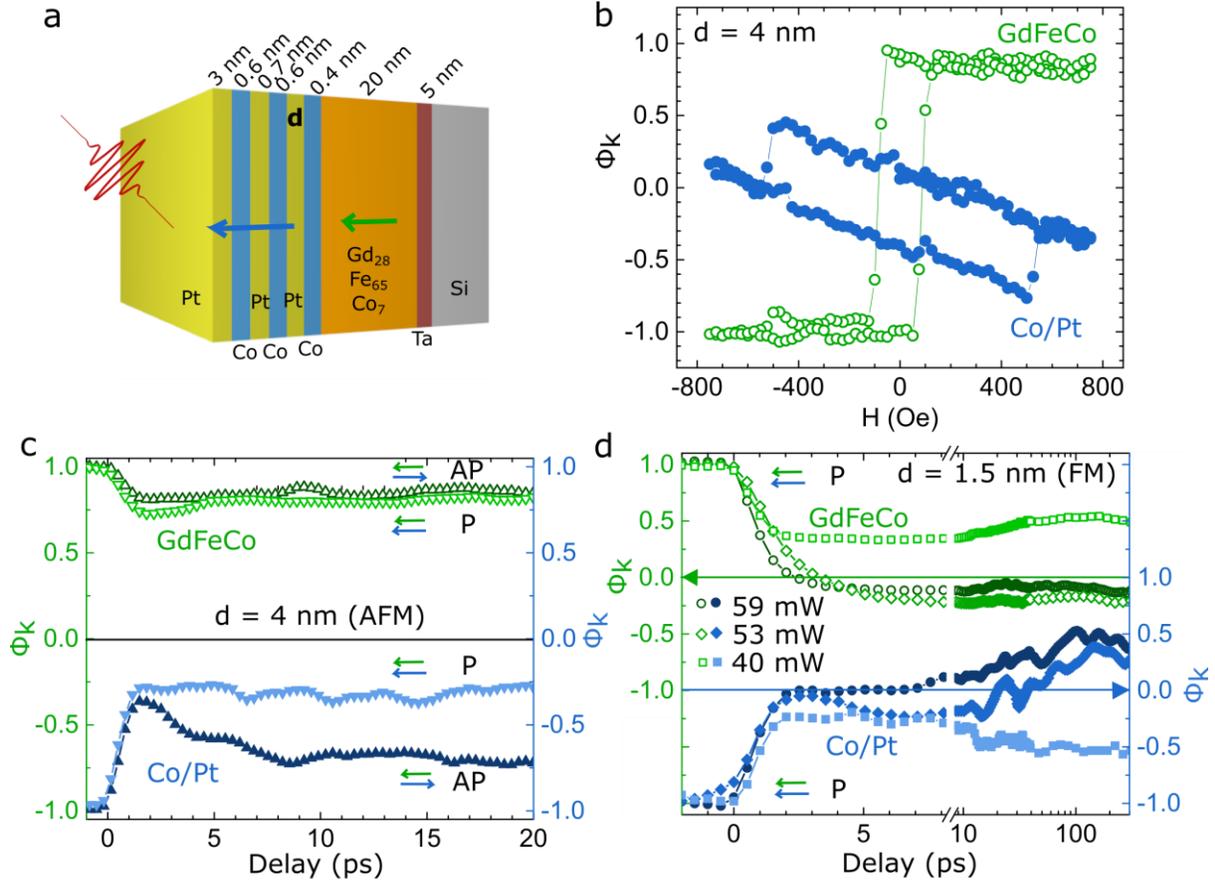

**Fig. 3**. **Time-resolved depth-sensitive magneto-optical measurements of laser induced dynamics. a,** GdFeCo/Co/Pt(d nm)/[Co/Pt]$_2$ stack series with Pt spacer of thickness d. **b,** Depth-sensitive MOKE magnetic hysteresis loops on sample d=4 nm. **c,** Depth-sensitive time-resolved demagnetization curves for antiparallel (AP) or parallel (P) initial states of the stack d=4 nm. Green and blue arrows represent GdFeCo and Co/Pt magnetizations, respectively. **d,** Depth-sensitive demagnetization and AOS experiments at various fluences on sample d=1.5 nm.